\documentclass[twocolumn,showpacs,preprintnumbers,amsmath,amssymb,prl]{revtex4}

\usepackage{graphicx}

\begin{document}

\title{Phase transitions towards frequency entrainment in large oscillator lattices}

\author{Per \"{O}stborn\thanks{Electronic mail: per.ostborn@matfys.lth.se}}
\author{Sven \r{A}berg}
\author{Gunnar Ohl\'en}

\affiliation{Division of Mathematical Physics, Lund University, S--221 00 Lund, Sweden}
        
\begin{abstract}
We investigate phase transitions towards frequency entrainment in large, locally coupled networks
of limit cycle oscillators. Specifically, we simulate two-dimensional lattices of pulse-coupled oscillators
with random natural frequencies, resembling pacemaker cells in the heart.
As coupling increases, the system seems to undergo two phase
transitions in the thermodynamic limit. At the first, the largest cluster of frequency entrained oscillators becomes
macroscopic. At the second, global entrainment settles. Between the two
transitions, the system has features indicating self-organized criticality.
\end{abstract}


\maketitle

Many systems in science, engineering and social life
can be described as large networks of coupled limit cycle
oscillators \cite{pikovsky}. Most often there is a spread in the individual, natural frequencies,
and the coupling is such that it tends to even out these frequency differences.
A general question is how the dynamics of such systems changes when the coupling between the oscillators increases.
Is there a phase transition at which the oscillators attain a common, collective frequency,
in the thermodynamic limit where the number $N$ of oscillators goes to infinity?

Such phase transitions are relevant in several fields.
For example, the proper function of the millions of pacemaker cells in the sinus node in the heart,
requires that they work at the same frequency.           
Cardiac arrhythmias may result if this is not the case \cite{sicksinus}.
The appearance of several frequencies in the sinus node may be caused by decoupling
due to tissue degeneration. The brain also contains many
pacemaker cells. Increasing evidence suggests that enhanced electric coupling between neighbor
neurons can provoke epileptic seizures \cite{carlen}. These correspond to pathologically large regions
of synchronized electric discharges. Thus, in this case a phase transition to a state of collective oscillation
is unfavorable, while it is vital for the coordinated function of the heart.
 
Most theoretical studies of such transitions assume that each oscillator is coupled to all the others
equally strong. The most well-known system of this kind is the Kuramoto model \cite{kurarev}.
More realistic networks have local coupling.
Winfree \cite{win1} and Kuramoto \cite{kura} hypothesized
that in systems with nearest neighbor coupling and random natural frequencies,
there should be a critical coupling strength at which the number of members,
or size $S_{max}$, of the largest cluster of frequency entrained \cite{ent} oscillators becomes macroscopic.
This can be expressed as a phase transition at which the order parameter $r$ becomes nonzero, where

\begin{equation}
r \equiv \lim_{N\rightarrow \infty} S_{max}/N.
\end{equation}

For a long time, model studies only revealed negative or inconclusive results regarding the existence
of such a phase transition \cite{prev}.
Recently, we proved that it was present in a one-dimensional chain
of pulse-coupled oscillators \cite{chain}. A finite critical coupling strength $g_{c}$ was found,
at which global frequency entrainment settles. At $g_{c}$, $r$ jumps discontinuously from zero to one.
In this Letter, a two-dimensional square lattice with bidirectional nearest neighbor coupling
is studied. Two phase transitions at the critical couplings $g_{c1}$ and $g_{c2}$ are found, at which
the order parameter $r$ seems to become non-zero and one, respectively.

\begin{figure}
\begin{center}
\includegraphics[clip=true]{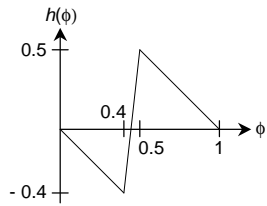}
\end{center}
\caption{The shown function $h(\phi)$ times the coupling constant $g$ is the PRC
         used in system (\ref{pulsesys}). From the requirement $0 \leq \phi +gh(\phi)<1$, we have $g<1$.
         As discussed in Ref. \cite{chain}, $g=1$ corresponds to infinite coupling.}
\label{prc}
\end{figure}

In our model (c.f. Ref. \cite{chain}), the state of oscillator $k$ is given by the phase $\phi_{k} \in [0,1)$.
The time evolution of the phase is given by
\begin{equation}
\dot{\phi}_{k}=1/P_{k}+g h(\phi _{k})\sum_{l\in n_{k}} \delta (\phi _{l}).
\label{pulsesys}
\end{equation}
$P_{k}$ is the natural period of oscillator $k$, and $n_{k}$ is the set of its nearest neighbors.
An oscillator $l$ is said to fire when $\phi_{l}=1$. Then $\phi_{l}\rightarrow 0$ and a pulse
is delivered to the neighbor $k$, so that its phase immediately shifts according to
$\phi_{k} \rightarrow \phi_{k}+g h(\phi_{k})$.
This kind of system can model oscillators that interact with short pulses and are strongly attracted to
their limit cycles. Examples include pacemaker cells
in the heart, neurons, flashing fireflies, chirping crickets, and people clapping their hands
in the theater.
The function $g h(\phi_{k})$ is called the phase response curve (PRC),
where $g$ is the coupling strength. Inspired by experiments on pacemaker cells in the heart \cite{sano},
we assume the form of the PRC given in Fig. \ref{prc}. This coupling tends to even out phase
and frequency differences between oscillators for the following reason:
If the phase $\phi_{k}$ of an oscillator $k$ receiving a pulse
from a neighbor $l$ is small, it becomes even smaller ($h<0$), approaching $\phi_{l}=0$. If $\phi_{k}$ is large,
it becomes even larger ($h>0$), again approaching $\phi_{l}$. We expect that it is this bipolar
character of the PRC that is essential for the observed dynamics, not its exact shape.

The natural periods $P_{k}$ are taken as random numbers from a square distribution with $P_{min}=1$ and
$P_{max}=1.5$ time units (t.u.). We use periodic boundary
conditions. The same method of numerical integration as in Ref. \cite{chain} is used.
The lattice is divided into blocks of
$10 \times 10$ oscillators, within which the integration is exact. These correspond to the segments
of $25$ oscillators within which the integration was exact in Ref. \cite{chain}.

\begin{figure}
\begin{center}
\includegraphics[clip=true,width=8cm]{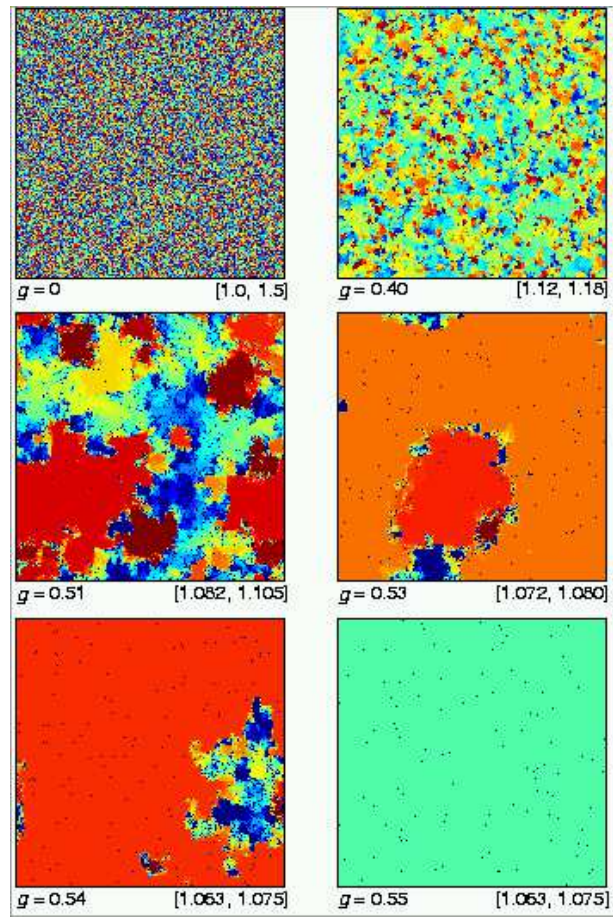}
\end{center}
\caption{Mean period landscapes in a lattice of $500\times 500$ oscillators for different coupling strengths $g$.
         The couplings $g=0$, and $g=0.40$ belong to phase 1
         where there are only microscopic frequency entrained clusters. At $g=0.51$ we are close to the
         critical value $g_{c1}$, where one cluster starts to dominate. The couplings $g=0.53$ and $g=0.54$ belong to phase 2,
         with one percolating, macroscopic cluster. At $g=0.55$ we have reached phase 3, where all oscillators
         frequency entrain, except for some silent oscillators (blue dots). The color codes for the mean period $P$
         in such a way that deep red corresponds to small $P\leq P_{-}$, and deep blue to large $P\geq P_{+}$.
         In each panel, the color scale is given as $[P_{-},P_{+}]$.}
\label{meanp1}
\end{figure}

The two phase transitions separate three phases. We shall say that states with $r=0$ belong
to ``phase 1'', states with $0<r<1$ to ``phase 2'', and states with $r=1$ to ``phase 3''. 
Figure \ref{meanp1} shows mean periods in a large lattice measured during $10^{4}$ t.u. after a transient
of $10^{5}$ t.u. for different coupling strengths $g$. For $g=0$ the mean periods are the natural
periods, which are independent random numbers. As we increase $g$, clusters of oscillators with nearly
identical mean periods appear, as can be seen for $g=0.40$.
The typical size of these clusters increases with $g$.
At $g=0.51$, one cluster is almost percolating
horizontally, indicating that this coupling is close to the critical value $g_{c1}$, which separates
phase 1 from phase 2. At $g=0.53$
and $g=0.54$, one cluster is percolating through the lattice. This cluster can be interpreted as
macroscopic, suggesting that we have entered phase 2.
For $g=0.55$ the entire lattice attains the same frequency, suggesting that we have entered phase 3.

Around $g=0.48$, oscillators that do not fire during the measurement interval start to appear.
They become more frequent as $g$ increases towards $g_{c2}$. However, their number decreases with increasing
measurement interval, showing that they are not silent forever. In phase 3, however,
there is a small number of oscillators ($<0.05\%$) that never fire. The others always fire with a
common interval.
Evidently, the silent oscillators are repeatedly perturbed by their neighbors at a phase $\phi$ where
the PRC is negative, so that the phase is kept in this interval, and does not reach the 
threshold $\phi=1$. This situation can only be maintained at all times if the surrounding is stable,
as it is in phase 3. To say that $r=1$ in phase 3, we have to exclude the silent oscillators.

\begin{figure}
\begin{center}
\includegraphics[clip=true]{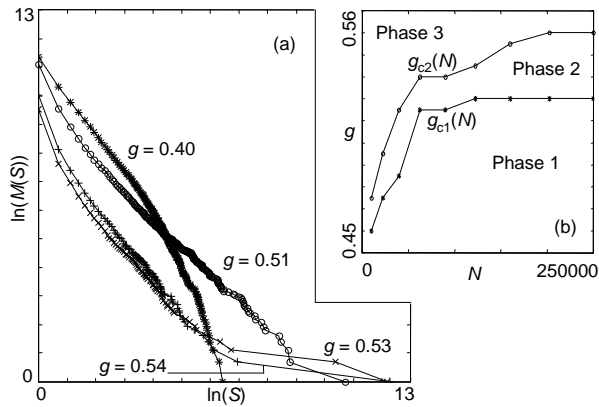}
\end{center}
\caption{(a) The number of clusters $M(S)$ with size equal to or larger than $S$ as a function of $S$ for different
         coupling strengths $g$ in double logarithmic scale. The data are taken from the systems shown in
         Fig. \ref{meanp1}. The distribution is sub-critical for $g=0.4$, approximately
         critical for $g=0.51$, and close to critical also for $g=0.53$ and $g=0.54$, if the largest
         clusters are disregarded. (b) Estimations of $g_{c1}(N)$ and $g_{c2}(N)$ with accuracy
         $\Delta g=0.005$, using a single realization.}
\label{clustersizes}
\end{figure}

To locate $g_{c1}$ more precisely, we study how the distribution of cluster sizes depends on $g$.
A cluster is numerically defined to be a connected set of oscillators whose mean periods do not differ
more than $dP=0.001$. The results are shown in Fig. \ref{clustersizes}(a). The distribution becomes critical
(i.e. a power law) at $g\approx 0.51$, suggesting that $g_{c1}$ is close to $0.51$.
At $g=0.53$ and $g=0.54$ the cluster
sizes are still close to critically distributed, if we exclude the largest clusters, which are too large
to fit in. For $g=0.53$ there are two such clusters, and for $g=0.54$ there is one. Thus
the system might be described as critical in the entire phase 2.
A cluster percolating through the lattice is not always seen in phase 2. However,
the largest clusters are always larger than they would be if all cluster sizes were critically
distributed. In some cases, a percolating cluster develops if the simulation is allowed to proceed
further. The opposite is never seen, i.e. that a percolating cluster disappears,
suggesting that a macroscopic, percolating cluster always appears sooner or later in phase 2.
The critical distribution for the smaller clusters is seen in all simulations in phase 2,
and it is robust with respect to the cluster discrimination parameter $dP$. Self-organized criticality
has been observed previously in
lattices of pulse-coupled oscillators with diverse natural frequencies \cite{corral}, where it appeared as
critically distributed avalanches of simultaneous firings. The question of
frequency entrainment was not addressed in that study. In our model, such avalanches are, however, impossible,
since waves of firings propagate with finite speed whenever $g<1$.

To confirm the existence of $g_{c1}$ and $g_{c2}$, one should ideally 
determine their magnitude as a function of $N$, and see whether they converge
to finite, separate values as $N\rightarrow\infty$.
We have not been able to execute this scheme fully, because of the long computation times
involved. Figure \ref{clustersizes}(b) shows estimations of $g_{c1}(N)$ and $g_{c2}(N)$
with accuracy $\Delta g=0.005$, using a single realization, meaning that a single assignment
of natural periods and a single initial condition (phases at time zero) is used for a given $N$. 
More realizations indicated that the spread of the critical values is similar to $\Delta g$
for $N=500^{2}$, i.e. considerably less than $g_{c2}(N)-g_{c1}(N)$.
The shape of the curves in Fig. \ref{clustersizes}(b) support the hypothesis
that the critical couplings converge to separate
finite values. Since a critical coupling $g_{c}=\sqrt{2/3}\approx 0.82$ for global frequency entrainment
was shown to exist
in the corresponding one-dimensional lattice \cite{chain}, we are strongly inclined to believe that a finite
$g_{c2}<0.82$ exists here, since increased connectivity in general facilitates
the appearance of order. Therefore the most important observation in Fig. \ref{clustersizes}(b) is that
$g_{c2}(N)-g_{c1}(N)$ does
not seem to decrease as $N$ increases, suggesting that $g_{c1}$ is indeed lower that $g_{c2}$.
To be confident that phase 2 between $g_{c1}$ and $g_{c2}$ exists, one should also check that
the lattice does not converge to a frequency entrained state in this coupling interval,
albeit much slower than above the presupposed larger $g_{c2}$. This was done for a $300\times 300$ lattice
at $g=0.525$, which was simulated during $4.4\times10^{5}$ t.u. After a transient of about
$1.5\times10^{5}$ t.u., the standard deviation of the mean period distribution in the lattice
did not show any tendency to decrease.

\begin{figure}
\begin{center}
\includegraphics[clip=true,width=8cm]{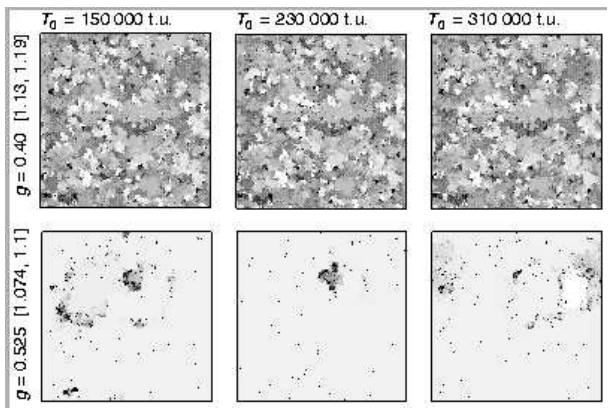}
\end{center}
\caption{Evolution of mean period landscapes in a lattice of $300\times 300$ oscillators for $g=0.40$
         (top row), belonging to phase 1, and $g=0.525$ (bottom row), belonging to phase 2.
         The clusters seem stable in phase 1 and unstable in phase 2. The mean periods were measured during
         $10^{4}$ t.u., starting from different times $T_{0}$.
         The color scale is given in square brackets in the same way as in Fig. \ref{meanp1},
         but here white corresponds to small $P$, and black to large $P$.}
\label{moving}
\end{figure}

Mean period landscapes from this simulation (in phase 2) are shown in the bottom row of Fig. \ref{moving}.
The cluster
configuration never seems to stabilize. However, some clusters exist for a long time. For example,
the white cluster in the leftmost panel exists to the end of the simulation, i.e at least
for $1.3\times 10^{5}$ t.u. These features of phase 2 are consistent
with the hypothesis of criticality. Then the system is expected to be self-similar in time,
without a characteristic time interval, during which mean periods could be calculated with confidence.
It should also lack a characteristic life-time for the clusters, so that some clusters exist
during short time intervals, and a few for very long times.
The upper row of Fig. \ref{moving} shows mean periods in the same lattice for $g=0.40$ (in phase 1),
calculated at corresponding times.
It is seen that the positions of the clusters remain essentially fixed. This seems always
to be the case in phase 1.
A related instability in phase 2 is that the mean period landscapes at a given time from
two simulations with different initial conditions look very different.
This would make the landscape in phase 2 sensitive to external perturbations.
In contrast, in phases 1 and 3 the system
seems to approach the same mean period landscape regardless the initial condition.
These instabilities made it impossible to determine if, and how, the order parameter $r$
changes with $g$ in phase 2.

The mean period of an oscillator is defined as time goes to infinity.
The instability of the clusters in phase 2 indicates that
these numbers, if they exist, could be different from mean periods that are measured during a finite time. 
We can imagine three possibilities: 1) The mean periods do not exist, 2) they exist and are
the same for all oscillators, and 3) they exist, and are different.
If frequency clusters appear and disappear at completely random
places in the network as time passes, then alternative two could be true. If not, we could have alternative three.
If this alternative is correct,
one could ask how the mean period landscape would look like. It is not self-evident that there would
still be a macroscopic, percolating cluster. If this is the case, or alternative 1) or 2) is true,
one has to assume a finite measurement time to say that $0<r<1$ in phase 2.

\begin{figure}
\begin{center}
\includegraphics[clip=true]{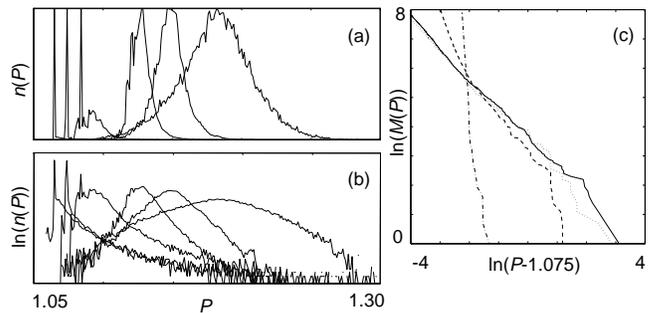}
\end{center}
\caption{Distributions $n(P)$ of mean periods $P$ for a single realization of a lattice of size $500\times 500$.
         (a) Linear scale, normalized heights.
         The rightmost peak corresponds to $g=0.30$, and going to the left we have $g=0.40$, $0.45$,
         $0.51$, $0.53$, and $0.54$. (b) Corresponding distributions in logarithmic scale.
         (c) The long period tails in double logarithmic scale. $M(P)$ is the number of oscillators
         with mean period equal to or larger than $P$.
         The dash-dotted line corresponds to $g=0.40$, the dashed to $g=0.51$, the dotted to $g=0.53$,
         and the solid to $g=0.54$. For the two latter couplings in phase 2, the tails follow a
         power law.}
\label{distri}
\end{figure}

In addition to the study of mean period landscapes, we investigate the distribution $n(P)$ of the mean periods $P$.
Figure \ref{distri} shows such distributions for different values of $g$.
It is seen that the maximum shifts towards smaller values of $P$
as $g$ increases. For a one-dimensional chain, it was shown that the entrained period was
always that of the fastest oscillator in the thermodynamic limit \cite{chain}. The same could very well
be true here, in which case the distribution would become a delta spike $\delta(P-1)$ at $g_{c2}$.
For $g\leq 0.4$, the distribution is approximately
symmetric around the maximum at $P_{0}$. Assuming a functional form $n(P)\propto \exp(-\alpha|P-P_{0}|^{\beta})$,
it seems that $\beta$ decreases through one as $g$ increases. Above $g=0.45$, the distribution becomes
progressively asymmetric, with a wider and wider tail of long periods.
It is seen in Fig. \ref{distri}(c) that this tail obeys a power law $n(P)\propto (P-P_{0})^{-\gamma}$ in phase 2.
Subtracting one from the slope in this cumulative plot, we get $\gamma\approx 2$.

The appearance of oscillators with long mean periods parallels the appearance of completely silent
oscillators, mentioned previously. To investigate the behavior of such slow or silent oscillators,
time series of firing intervals for individual oscillators are studied. It is seen that oscillators
with very long mean periods most often fire with normal intervals, but also experience very long periods
of silence, appearing as intermittent ``bursts''. Oscillators with more
normal mean periods also display intermittency, but the bursts of silence are much shorter.
The bursts become more scarce
as the coupling increases towards $g_{c2}$, as usual for an intermittent signal
as we approach the bifurcation point where they disappear.

In summary, we have found strong indications for the existence of two phase transitions in a large lattice
of pulse-coupled oscillators with diverse natural frequencies.
At the first transition, one frequency entrained cluster becomes macroscopic. At the second,
all oscillators frequency entrain. Between the two transitions, the system seems critical, with
self-similar cluster size distribution for the microscopic clusters, and strong fluctuations,
with possible self-similarity in time.
These phenomena calls for further investigations, in particular to find out
if they are generic for large locally coupled oscillator networks.

We thank Martin Folkesson for making the initial simulations in this study.


\begin{thebibliography}{10}
\bibitem{ent}That two oscillators are frequency entrained means that they have identical mean frequencies,
                 if these are measured during infinitely long time.
\end{thebibliography}
\end{document}